\documentclass[letterpaper, 10 pt, conference]{ieeeconf}

\IEEEoverridecommandlockouts

\newcommand{\RNum}[1]{\uppercase\expandafter{\romannumeral #1\relax}}

\usepackage{graphicx,wrapfig,fullpage,amsmath,hhline,epsfig,verbatim,url,amssymb,multicol,multirow,cite}
\usepackage{times,color,soul} 
\usepackage[left=0.75in,top=0.75in,right=0.75in,bottom=0.75in]{geometry}
\usepackage{amsbsy,latexsym,mathrsfs,mathtools}
\usepackage{tikz}
\usetikzlibrary{shapes.geometric, arrows}
\usepackage{mathptmx} 
\DeclareMathAlphabet{\mathcal}{OMS}{cmsy}{m}{n}
\usepackage{bm}
\usepackage{float}
\usepackage{color,soul}
\usepackage{dsfont}
\usepackage{comment}
\usepackage{algorithm}
\usepackage{algorithmic}
\usepackage{psfrag}  

\usepackage{flushend}

\usepackage{url}
\usepackage{hyperref}
\usepackage{setspace}
\usepackage[utf8]{inputenc}
\usepackage{array}
\newcolumntype{P}[1]{>{\centering\arraybackslash}p{#1}}

\title{\LARGE \bf
Modeling of Interface Loads for EOD Suit Wearers}

\author{Yuan Gao$^{1}$, Stephanie Epstein$^{1}$, Murat Inalpolat$^{1}$, Yi-Ning Wu$^{2}$, and Yan Gu$^{3}$
\thanks{$^{1}$Y. Gao, S. Epstein, and M. Inalpolat are with the Department of Mechanical Engineering, 
and $^{2}$Y.-N. Wu is with the Department of Physical Therapy and Kinesiology,
University of Massachusetts Lowell, Lowell, MA 01854, U.S.A.
E-mails: {\{\tt \small yuan\_gao, stephanie\_epstein\}@student.uml.edu}, {\tt \small \{murat\_inalpolat, yining\_wu@uml.edu}.
$^{3}$Y. Gu is with the School of Mechanical Engineering, Purdue University, West Lafayette, IN 47907, U.S.A.
    E-mail: {\tt \small yangu@purdue.edu}.}
}
\begin{document}
	\maketitle
	\thispagestyle{empty}
	\pagestyle{plain}

\begin{abstract}
Explosive Ordnance Disposal (EOD) suits are widely used to protect human operators to execute emergency tasks such as bomb disposal and neutralization.
Current suit designs still need to be improved in terms of wearer comfort, which can be assessed based on the interaction forces at the human-suit contact regions.
This paper introduces a simulation-based modeling framework that computes the interaction loads at the human-suit interface based on a wearer's kinematic movement data.
The proposed modeling framework consists of three primary components:
a) inertial and geometric modeling of the EOD suit, b) state estimation of the wearer's in-suit movement, and c) inverse dynamics analysis to calculate the human-suit interface forces based on the simulated human-suit model and the estimated human movement data.
This simulation-based modeling method could be used to complement experimental testing for improving the time and cost efficiency of EOD suit evaluation.
The accuracy of the simulated interface load was experimentally benchmarked during three different human tasks (each with three trials), by comparing the predicted interface forces with that measured by commercial pressure sensors.

\end{abstract}

\section{INTRODUCTION}

Various capabilities of the existing explosive ordnance disposal (EOD) suits have been extensively studied~\cite{dale2005methodology,hayda2004blast,gmitrzuk2018influence,dionne2019investigating}, with a primary focus on blast and heat protection.
In contrast, only a few studies have investigated the ergonomics of existing EOD suits~\cite{hennessy2006results,deane2021pressure} in terms of user comfort and fatigue. 
Yet, the ergonomics of other full-body, heavy-weight, protective suits, such as the Extravehicular Mobility Units (EMUs), have been extensively studied with a focus on the physical suit-human interaction that can be used to indicate user comfort.
These studies have revealed that existing EMU designs (e.g., space suits) could cause user discomfort by inducing injuries and significantly boosting wearers' metabolic costs~\cite{chesterton2003gender,williams2003emu,strauss2004extravehicular}.
These negative effects may compromise the operational performance of a suit wearer during task execution~\cite{Anderson2014UnderstandingHS}.
Thus, it is essential to quantify the physical human-suit physical interaction for users wearing full-body, heavy-weight, protective suits that include both EMUs and EOD suits.

The physical interaction between a wearer and a space suit has been recently investigated.
Diaz and Newman have proposed an approach to measure the physical human-suit interaction as well as the joint torque~\cite{6836247}, by modelling the interaction forces as an external load applied to the human subject.
Yet,
modelling the human-suit interaction as a pre-specified external load applied at a point may not accurately reflect the interface load because the interaction typically occurs within a finite region instead of a point.

To accurately capture the physical interaction at the human-suit interface, a pressure sensing system has been developed to experimentally measure the interface loads between the human and suit~\cite{anderson2015development,anderson2015pressure,hilbert2015human,7500699}.
To further investigate the interaction between the space-suit and wearer, a sensing system with additional capabilities (e.g., temperature and humidity sensing) has been developed~\cite{shen2018wearable}. 

Although pressure sensing systems could be used to directly measure the interface load experienced by EOD suit wearers, experimental pressure sensing during various movements of wearers could be time-consuming (e.g., due to the time costs of the calibration, placement, and re-zeroing of pressure sensors~\cite{deane2021pressure}).
To this end, simulation-based modelling could be exploited to compute the interface loads without utilizing experimental pressure sensing, thus complementing experimental testing and alleviating the burden of extensive tests.
In this study, we introduce a simulation-based modeling framework that uses biomechanics simulation software to calculate the interaction forces between the wearer and the EOD suit during different full-body motions.
The framework includes an integrated human-suit model that captures the realistic human biomechanics, the essential features of the inertial and geometrical properties of the EOD suit, and the physical interaction between the suit and the human model within the finite contact regions.
Based on the integrated human-suit model, the framework also incorporates inverse dynamics analysis, which is performed via biomechanics software, to compute the reaction forces at a set of user-defined contact regions and points based on the wearer's movement data. 
The main contributions of this work are: (a) proposing a new method to obtain the pressure data between the wearer and suit with various motions rather than relying on the human subject experiments solely and (b) emulating the suit-human interactions using rigid bodies and various constraints.
Results of pilot experiments validated the effectiveness of the framework in modeling the wearer-suit interface loads during different mobility tasks.

\begin{figure}[t]
    \centering
    \includegraphics[width=0.75\linewidth]{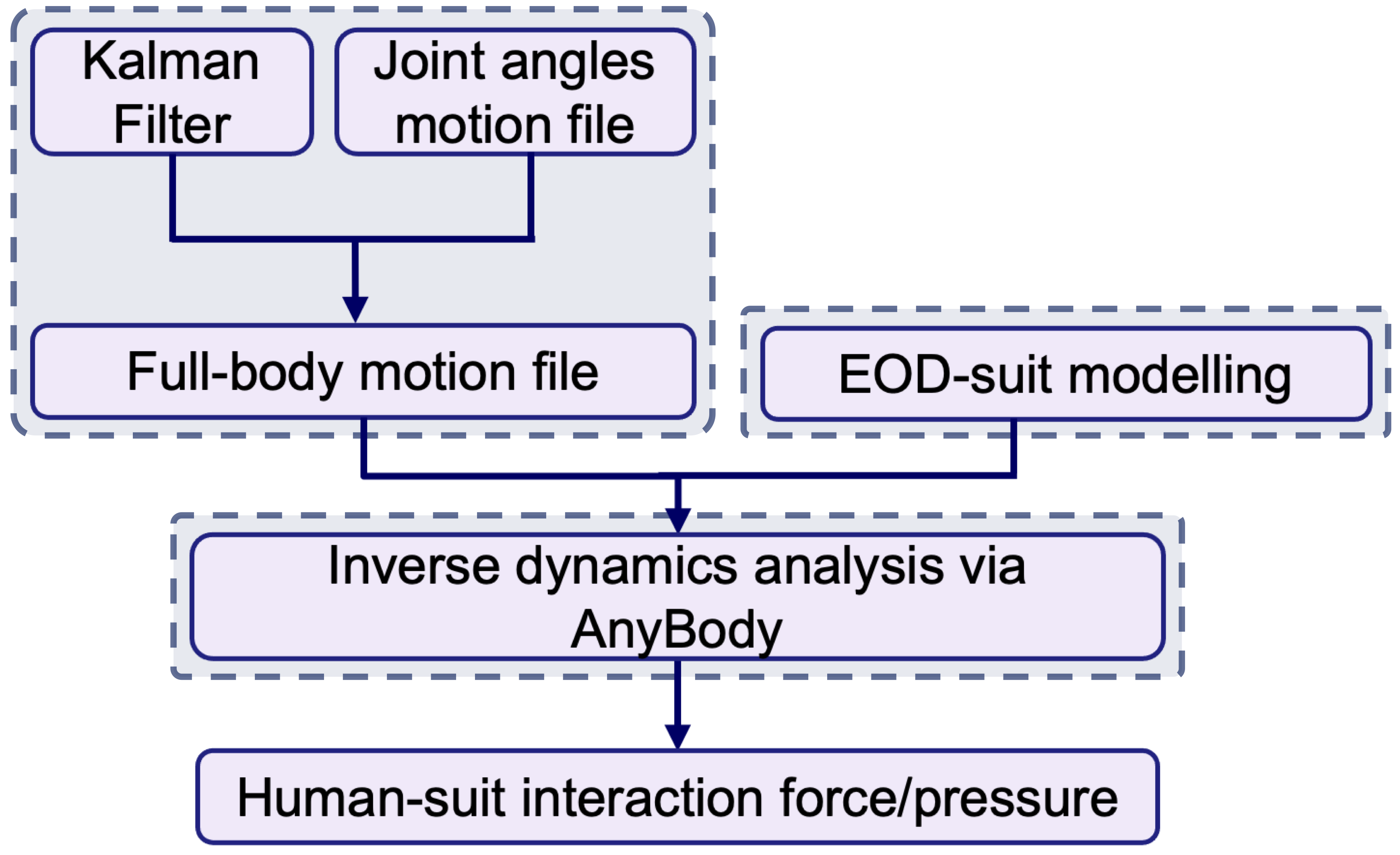}
    \vspace{-0.1 in}
    \caption{Overview of the proposed modeling framework that comprises three main components (highlighted with dashed blocks).
    The three components are suit modeling, in-suit motion estimation, and inverse dynamics analysis.}
    \label{fig:validation overview}
\end{figure}

\section{SIMULATION-BASED HUMAN-SUIT MODELING}
\label{sec:model and method}

This section presents the proposed simulation-based approach of human-suit modeling.
The objective of the modeling is to accurately produce the interface loads between the wearer and the suit based on the wearer's movement data.

To reach the modeling objective, the proposed approach comprises three main components (see Fig.~\ref{fig:validation overview}).
The first component is the modeling of the EOD suit to capture its essential physical properties (e.g., mass and geometry) that could affect the human-suit interface loads at the critical regions (e.g., shoulders), as introduced in subsection A. 
The geometry modeling is performed in SOLIDWORKS.

The second component is the human movement estimation to obtain the kinematic data (e.g., the global position of the wearer) that is needed to compute the interface loads using biomechanics software but cannot be directly measured, as explained in subsection B.
Note that the human movement data is required for interface load computation since a wearer's movement can directly affect the interface loads.

The last component is the inverse dynamics analysis via biomechanics-based simulation for calculating the interface loads based on the outcomes from the first two components (i.e., integrated human-suit model and the estimated human movement), as presented in subsection C.
We choose to use physics-based simulations, instead of analytical methods (e.g., mathematically modeling the interaction based on physics laws), as the basis to study the interface loads.
This is due to the fact that the human-suit interaction is complex, involving contact areas at multiple locations, complex geometry of both the human and the suit, and different load patterns under different human motions.
In other words, it may not be tractable to model the physical interface loads using analytical methods.

\subsection{EOD Suit Modeling}

This subsection introduces the proposed modeling of the EOD suit in SOLIDWORKS to capture the essential inertial and geometric properties of the suit.
The suit model created is integrated with a high fidelity human model in biomechanics software for interface force computation as explained in subsection C.

\begin{figure}[t]
    \centering
    \includegraphics[width=1\linewidth]{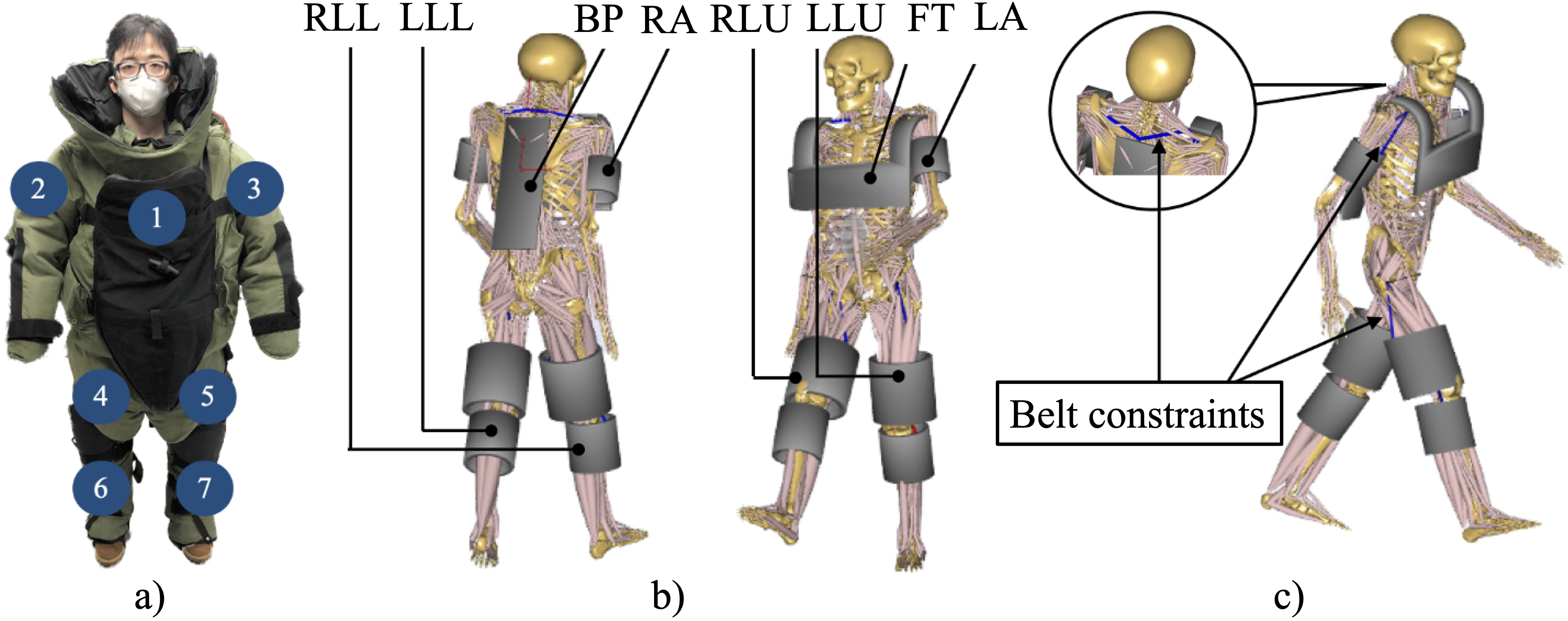}
    \vspace{-0.2 in}
    \caption{Illustrations of a) the tested subject wearing the EOD 8 Suit, b) eight major components of the proposed suit model, which are created in SOLIDWORKS and assembled to the human model in AnyBody, and c) ``belt'' constraints anchoring the suit components to the human model in AnyBody. The labels in subplot a) highlight seven of the eight components: (1) FT; (2) RA; (3) LA; (4) RLU; (5) LLU; (6) RLL; and (7) LLL.}
    \label{fig:validation of the human-suit interaction setup2}
\end{figure}

The EOD suit of interest to this study is the ``EOD 8 Suit'' (see Fig.~\ref{fig:validation of the human-suit interaction setup2} a)which is a heavy, full-body suit designed to protect the wearer from the heat and shockwaves induced by a bomb or any fragments the bomb may generate.
The EOD 8 suit has been in service since 1999 and is one of the most widely used EOD suits for bomb disposal operations around the world~\cite{dale2005methodology,deane2021pressure}.

The EOD 8 Suit utilized in this study is medium-small sized, and the total mass of its main components (without the helmet and the groin portion) is approximately 18.25 kg (i.e., 179.01 N).
Its outer fabric is made of an aramid weave, within which alloy plates are installed at the chest, back, knee portions for providing additional protection.

The inertial and geometrical properties of the EOD suit are complex because the suit comprises multiple rigid (e.g., metal pads inserted within the suit) and soft pieces (e.g., fabrics) with complex shapes.
To provide a relatively accurate representation and model of the EOD suit for efficient interface load computation, we use SOLIDWORKS to build a simplified three-dimensional (3-D) model of the EOD suit that captures the essential features of the suit such as its inertia and geometry.

\subsubsection{Modeling Assumptions}
The following model simplifying assumptions are considered:
\begin{itemize}
    \item[(A1)] The suit is modeled as a collection of rigid bodies.
    \item[(A2)] The density of each segment of the suit model is assumed to be evenly distributed.
    \item[(A3)] The helmet and the soft armor at the groin portion is omitted from the suit model.
\end{itemize}

Assumptions (A1) and (A2) are reasonable because the majority of an EOD suit's weight is contributed by the lumped alloy plates located at the subject's chest, back, and knees and the density of these alloy plates is evenly distributed.
Assumption (A3) is mainly for simplifying the suit modeling, and the helmet and the soft groin armor will be considered in our future work of suit modeling.

\subsubsection{Suit Component Modeling}
Under assumptions (A1)-(A3), we decompose the EOD suit (without the helmet and the groin portion) into the following eight parts (Fig.~\ref{fig:validation of the human-suit interaction setup2}-a):
(1) Right Leg Upper (RLU);
(2) Right Leg Lower (RLL);
(3) Left Leg Upper (LLU);
(4) Left Leg Lower (LLL);
(5) Back Pad (BP);
(6) Right Arm (RA);
(7) Left Arm (LA); and
(8) Front Torso (FT).

We model the shape and dimensions of each component in SOLIDWORKS based on those of a representative suit wearer, the EOD suit, and the wearer-suit contact region.
The eight suit components are illustrated in Fig.~\ref{fig:validation of the human-suit interaction setup2}-b.
Note that the model of the FT component has both shoulder and chest parts but does not include an abdominal part because the most significant pressure for the subject's upper body are at the shoulders and the chest~\cite{deane2021pressure}.

To obtain the precise weight of each major component of the EOD suit (specifically for the version EOD 8), we measured each suit component using a force plate for ten times and used the average value to represent its weight.
The force plate is the BMS600900 platform developed by Applied Molecular Transport Inc., with an accuracy of 0.05\% of the load and a resolution of 0.169 N.
The measured weight of the suit components is listed in Table~\ref{tab: suit weight}.

\begin{table}[t]
\centering
\caption{\small{Weight of the major components of the EOD suit}} 
\small
\begin{tabular}{ P{4cm}|P{1.5cm} }
\hline
\hline
\centering
Major component & Weight (N) \\
\hline
 \centering Right Leg Upper (RLU) & 9.34 \\
 \centering Right Leg Lower (RLL) & 14.01 \\
 \centering Left Leg Upper (LLU) & 9.28 \\
 \centering Left Leg Lower (LLL) & 13.92 \\
 \centering Back Pad (BP) & 12.21 \\
 \centering Right Arm (RA) & 13.51 \\
 \centering Left Arm (LA) & 13.51 \\
 \centering Front Torso (FT) & 93.23 \\
\hline
\end{tabular}
\label{tab: suit weight}
\end{table}

With the individual segments modeled in SOLIDWORKS, we then import the individual segments into biomechanics software to assemble the suit components and the human body (see Fig.~\ref{fig:validation of the human-suit interaction setup2}-b, c), as explained in subsection C.

\subsection{In-Suit Kinematics Measurement and Estimation}

Movement data is required by the inverse dynamics analysis. However, it cannot be directly measured based on the raw data returned by common sensors. Wearerble sensors, such as APDM~\cite{APDM}, estimate the joint angle of human subject.
Yet, they do not return the global position.
Therefore, state estimation methods are needed to produce the subject's global position based on movement data returned by wearerable sensors.

\subsubsection{Movement Sensors Selected}

\begin{figure}[t]
    \centering
    \includegraphics[width=0.9\linewidth]{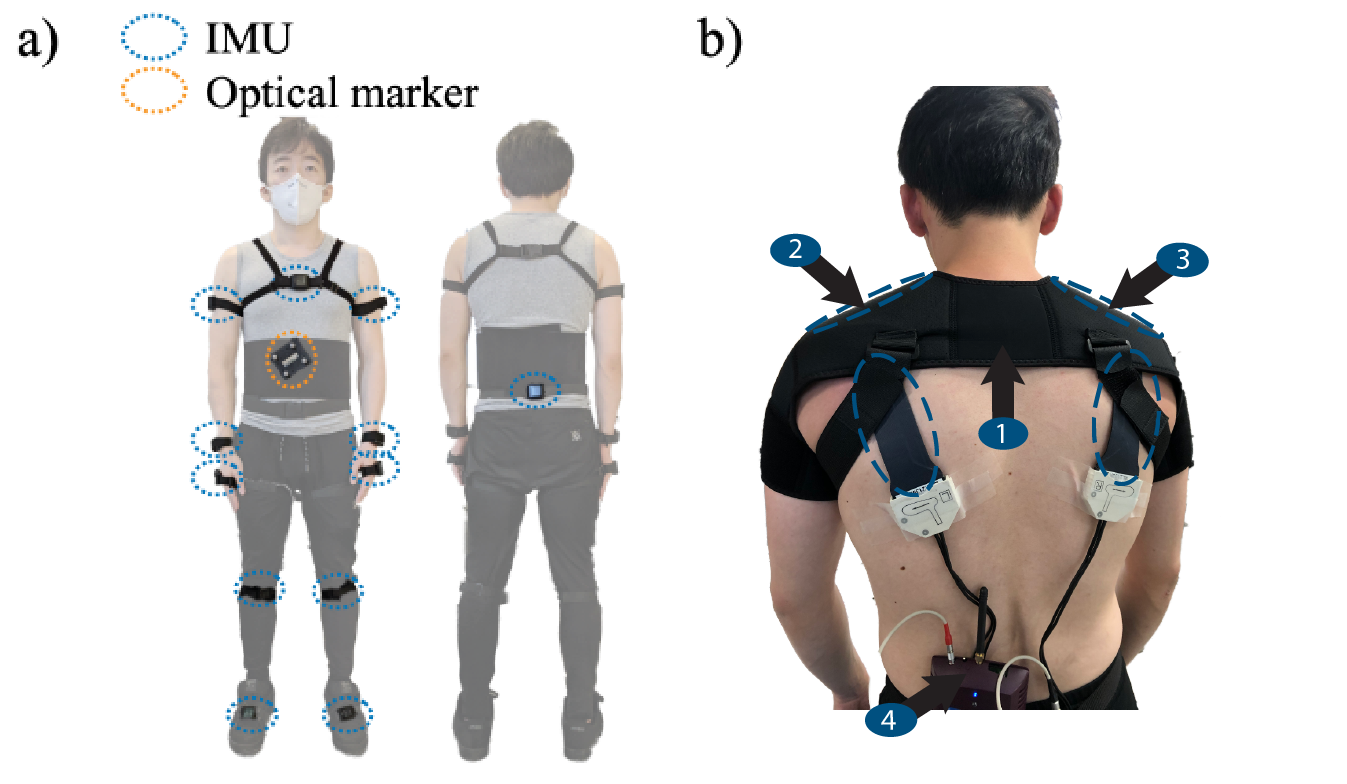}
    \vspace{-0.1 in}
    \caption{Illustrations of sensor placement: 
    a) IMU placement on the whole body of the human subject.
    The IMUs are used to obtain the joint angles of the subject during suited movement. 
    b) Pressure sensor placement at the subject's shoulders, with
    (1) shoulder straps used to secure the pressure sensor pads;
    (2) pressure sensor on the left shoulder;
    (3) pressure sensor on the right shoulder; and
    (4) Bluetooth wireless data transmitter.}
    \label{fig:IMUs placement}
\end{figure}

To reduce the discomfort caused by placing sensors on a suit wearer, we choose to use inertial-based motion capture systems that are compact and lightweight (see Fig.~\ref{fig:IMUs placement}-a).

The sensing system we use is the APDM~\cite{APDM} inertial motion-capture system, which provides the joint angles and the 3-D orientation of each IMU in the world.
As the the global position of the human subject is not directly returned by APDM but is often needed by inverse dynamics analysis (e.g., via AnyBody),
we choose to develop a state estimator based on Kalman filtering to obtain the global position data.
To that end, besides IMUs that are attached to each body limb for measuring the joint angles, we use the IMU placed at the lower back (i.e., base) to directly measure the linear acceleration and angular velocity of the base with respect to the IMU frame. 
The method used in this section can be found at~\cite{bledt2018cheetah}. In the following parts, process model and measured model will be introduced. The computational detail of Kalman filter is omitted due to the space limitation.

\subsubsection{Estimated Movement Variables}
\begin{figure}
    \centering
    \includegraphics[width=1\linewidth]{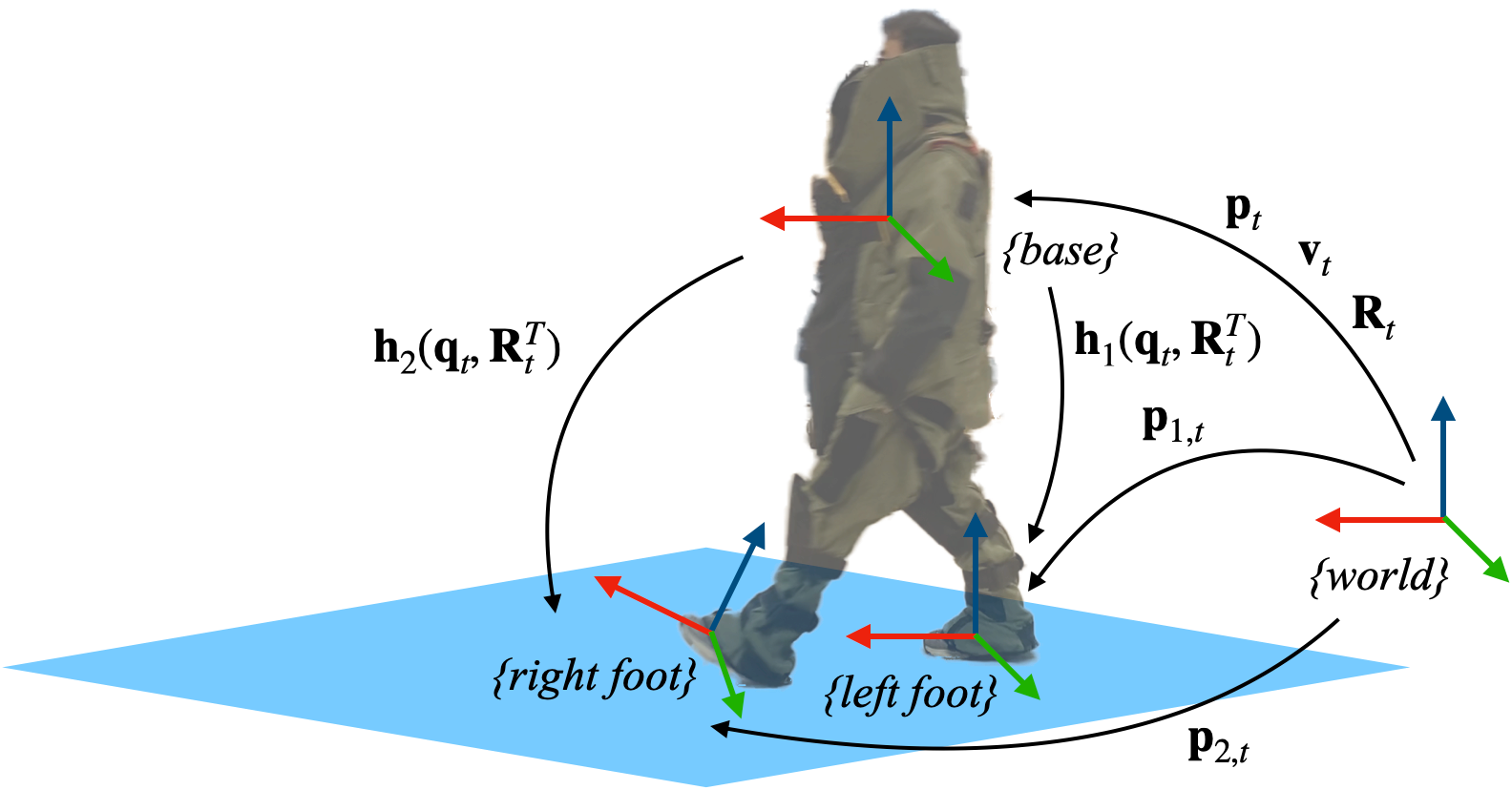}
    \vspace{-0.2 in}
    \caption{Illustration of the position and orientation variables and reference frames used in the proposed Kalman filter.
    The reference frame {\it \{world\}} is the world frame.
    The reference frames {\it \{right foot\}}, {\it \{left foot\}}, and {\it \{base\}} are attached to the subject's right foot, left foot, and base (i.e., lower back).}
    \label{fig:state illustration}
\end{figure}

The state of interest to be estimated is compactly expressed as
$\mathbf{x}_t = [\mathbf{p}_t^T,~\mathbf{v}_t^T,~\mathbf{p}_{1,t}^T,~\mathbf{p}_{2,t}^T]^T$,
where $\mathbf{p}_t \in \mathbb{R}^{3}$ is the base position in the world frame,
$\mathbf{v}_t \in \mathbb{R}^{3}$ is the base velocity in the world frame, and
$\mathbf{p}_{1,t} \in \mathbb{R}^{3}$ and $\mathbf{p}_{2,t} \in \mathbb{R}^{3}$ are left and right foot positions in the world frame.
Note that the subscript $t$ indicates the time instant $t$ and $(\cdot)_t$ denotes the value of the variable $(\cdot)_t$ at time $t$
All of these variables and reference frames are illustrated in Fig.~\ref{fig:state illustration}.

\subsubsection{Process Model}
As the APDM sensor system returns data at discrete times, the process model of the Kalman filter is designed in discrete time.
The filter design assumes that the IMU attached to the base gives sufficiently accurate data of the 3-D base orientation $\mathbf{R}_t\in \mathbb{R}^{3 \times 3}$ in the world frame.

Let the scalar variable $\Delta t$ be the duration between two successive sampling events.
Based on the dynamics of the base IMU~\cite{bloesch2013state,gao2021invariant,gao2022invariant}, the process model of the base position and velocity at time $t$ is given by:
\begin{equation}
\small
\label{equ: process model0-1}
    \begin{aligned}
    \mathbf{p}_{t+1} &= \mathbf{p}_{t}+\mathbf{v}_{t}\Delta t+\tfrac{\Delta t^2}{2}\mathbf{R}_t(\mathbf{y}_{a,t}+\mathbf{g});
    \\
    \mathbf{v}_{t+1} &= \mathbf{v}_{t}+\Delta t\mathbf{R}_t(\mathbf{y}_{a,t}+\mathbf{g}).
    \end{aligned}
\end{equation}
Here, the vector $\mathbf{y}_{a,t}\in \mathbb{R}^3$ is the accelerometer reading.
Then, $\mathbf{R}_t\mathbf{y}_{a,t}$ is the true value of the linear acceleration of the base IMU expressed in the world frame.

Based on the dynamics of the feet~\cite{bloesch2013state,zhu2022design,zhu2022invariant}, the process models of the left and right foot positions at time $t$ are:
\begin{equation}
\small
\label{equ: process model0}
    \begin{aligned}
    \mathbf{p}_{1,t+1} &= \mathbf{p}_{1,t}
    \quad
    \mbox{and}
    \quad
    \mathbf{p}_{2,t+1} &= \mathbf{p}_{2,t}.
    \end{aligned}
\end{equation}
Here, 
$\mathbf{p}_{1,t} \in\mathbb{R}^3$ and $\mathbf{p}_{2,t} \in\mathbb{R}^3$ are the positions of the left and the right feet expressed in the world frame, respectively.
If $\mathbf{p}_{i,t}~(i=1,2)$ is the stance foot position and the stance foot is static on the ground, then $\mathbf{p}_{i,t+1}=\mathbf{p}_{i,t}$ holds. 
However, if $\mathbf{p}_{i,t}$ is the swing foot position, then the process model $\mathbf{p}_{i,t+1} = \mathbf{p}_{i,t}$ no longer holds.
Accordingly, we set the covariance of this foot to be significantly large to effectively deactivate the process model of that foot position.

These process models can be compactly expressed as:
\begin{equation*}
\footnotesize
    \label{equ: process model mat}
    \begin{aligned}
    \underbrace{
    \begin{bmatrix}
    \mathbf{p}_{t+1}
    \\
    \mathbf{v}_{t+1}
    \\
    \mathbf{p}_{1,t+1}
    \\
    \mathbf{p}_{2,t+1}
    \end{bmatrix}
    }_{=:\mathbf{x}_{t+1}}
    =&
    \underbrace{
    \begin{bmatrix}
    \mathbf{I}_3 & \mathbf{I}_3\Delta t & \mathbf{0}_{3\times 3} & \mathbf{0}_{3\times 3}
    \\
    \mathbf{0}_{3\times 3} & \mathbf{I}_3 & \mathbf{0}_{3\times 3} & \mathbf{0}_{3\times 3}
    \\
    \mathbf{0}_{3\times 3} & \mathbf{0}_{3\times 3} & \mathbf{I}_{3} & \mathbf{0}_{3\times 3}
    \\
    \mathbf{0}_{3\times 3} & \mathbf{0}_{3\times 3} & \mathbf{0}_{3\times 3} & \mathbf{I}_{3}
    \end{bmatrix}
    }_{=:\mathbf{A}_t}
    \underbrace{
    \begin{bmatrix}
    \mathbf{p}_{t}
    \\
    \mathbf{v}_{t}
    \\
    \mathbf{p}_{1,t}
    \\
    \mathbf{p}_{2,t}
    \end{bmatrix}
    }_{=:\mathbf{x}_{t}}
    +
    \underbrace{
    \begin{bmatrix}
    \frac{\Delta t^2}{2}\mathbf{R}_t(\mathbf{y}_{a,t}+\mathbf{g})
    \\
    \Delta t(\mathbf{R}_t(\mathbf{y}_{a,t}+\mathbf{g})
    \\
    \mathbf{0}_{3\times 1}
    \\
    \mathbf{0}_{3\times 1}
    \end{bmatrix}
    }_{=:\mathbf{v}_{t}},
    \end{aligned}
\end{equation*}
where $\mathbf{I}_3$ and $\mathbf{0}_{3 \times 3}$ are $3 \times 3$ identity and zero matrices.

\subsubsection{Measurement Model}
When the sensors return data at time $t$, the update step of the KF is performed based on measurement models.
In this study, we form measurement models based on the forward kinematic chain connecting the base and the foot frames.
Let $\mathbf{h}_1 (\mathbf{{q}}_t)$
and
$\mathbf{h}_2  (\mathbf{{q}}_t)$
be the nonlinear forward kinematics functions representing the left and right foot positions with respective to the base frame, respectively.
Then, by the definition of $\mathbf{h}_i$ ($i=1,2$), we have $\mathbf{R}_{t}^T \mathbf{h}_i (\mathbf{{q}}_t) = \mathbf{p}_{i,t} - \mathbf{p}_t$.

Let the vector $\mathbf{q}_t$ be the wearer's joint angles obtained by APDM sensors at time $t$.

The measurement model of the filter is expressed as:
\begin{equation}
\small
\label{equ: measurement model}
\begin{aligned}
    \mathbf{R}_t^T \mathbf{h}_1({\mathbf{q}}_t) 
    =  \mathbf{p}_{1,t} - \mathbf{p}_t ~\mbox{and}~
    \mathbf{R}_t^T \mathbf{h}_2({\mathbf{q}}_t) 
    =  \mathbf{p}_{2,t} -\mathbf{p}_t.
\end{aligned}
\end{equation}

Equation~\eqref{equ: measurement model} can be organized into:
\begin{equation*}
\small
 \underbrace{
    \begin{bmatrix}
       \mathbf{R}_t^T \mathbf{h}_1(\mathbf{{q}}_t)
        \\
       \mathbf{R}_t^T \mathbf{h}_2(\mathbf{{q}}_t)
    \end{bmatrix}
    }_{=:\mathbf{h}}
    =
    \underbrace{
    \begin{bmatrix}
    \mathbf{I}_3 &\mathbf{0}_{3\times 3} & -\mathbf{I}_3 &\mathbf{0}_{3\times 3}
    \\
    \mathbf{I}_3 &\mathbf{0}_{3\times 3} & \mathbf{0}_{3\times 3} &-\mathbf{I}_{3}
    \end{bmatrix}
    }_{=:\mathbf{C}}
    \underbrace{
    \begin{bmatrix}
    \mathbf{p}_{t}
    \\
    \mathbf{v}_{t}
    \\
    \mathbf{p}_{1,t}
    \\
    \mathbf{p}_{2,t}
    \end{bmatrix}
    }_{=:\mathbf{x}_{t}}
\end{equation*}

\subsection{Interface Load Computation via Simulation-based Inverse Dynamics Analysis}
\label{subsec:Available Software Candidates Screening}

This subsection explains the computation of the human-suit interaction force based on simulation-based inverse dynamics analysis.
The analysis utilizes the previously explained suit model and the estimated human movement.

\subsubsection{Selection of Inverse Dynamics Analysis Software}

To reach our modeling objective of accurately producing the wearer-suit interface loads based on the wearer's movement data, the software should possess the following features.
First, the biomechanics model of the human should be reasonably accurate.
Second, the software should be capable of computing the contact force between the wearer and the suit in a realistic way, e.g., by explicitly considering the realistic physical interaction within the finite contact areas.
AnyBody software meets these requirements as it has high-fidelity customized human model. It also allows the computation of the subject-suit reaction force.

\subsubsection{Selection of Human Biomechanics Model in AnyBody}

The human model used here is a generic human body model provided by the AnyBody Managed Model Repository, which can be customized based on the actual subject's limb lengths, overall height, and weight.
In total, the human model in Anybody has 408 degrees of freedom and 214 joints.

The 3-D suit model created in SOLIDWORKS, as explained in subsection A, is a group of disconnected components corresponding to the eight major parts of a typical EOD suit.
We need to appropriately integrate the suit model with the realistic human model in AnyBody for computing the interface load based on the human's movement data, which is explained next.

\subsubsection{Contact Region Definition}

The suit and the human make contact at multiple finite sized regions, specifically, at infinitely many points within those contact regions.
Yet, computing the interface loads at infinitely many points may not be tractable.
To that end, we choose to simplify the interaction force computation by exploiting the built-in functionality of AnyBody that allows users to define a finite set of contact regions on both the suit and the subject for interaction force computation.
With AnyBody, each contact point within a contact region between the human model and an external object/environment is defined by: a) the position of the point in a 3-D Cartesian coordinate frame fixed to the suit and b) a local 3-D Cartesian coordinate system attached to the suit with its $y$-axis aligned with the normal direction of the contact surface at that point.

\subsubsection{``Belt'' Constraint Design}
To ensure that the eight components of the suit remain a secured contact with the human body, we use the ``belt'' constraint provided in AnyBody to anchor the suit components on the human body. 
Without the ``belt'' constraints enforced, the disconnected individual suit components will fall off the human body, and the simulator will report an error.
The belt only applies ``pulling'' forces between the connected suit and human segments, mimicking the suit's highly stiff fabric that connects different metal segments of the suit.
The belt can be defined by specifying its two end points, with one on the suit and the other on the body. 

We choose to set the belt constraints for different suit parts as follows (see Fig.~\ref{fig:validation overview} b):
\begin{enumerate}
    \item [(a)] Back Pad (BP) and Front Torso (FT) are connected to a single point on the lower part of the neck.
    \item [(b)] Each Upper Leg (LLU or RLU) is connected to a single point on the outer side of the hip.
    \item [(c)] Each Lower Leg (LLL or RLL) is connected to a single points on the outer side of the knee.
\end{enumerate}

\subsubsection{Interface Load Computation via AnyBody Inverse Dynamics Analysis}

After setting up the integrated human-suit model in AnyBody,
the inverse dynamic analysis can be performed to obtain the 3-D reaction force at each contact point.
These forces can then be used to compute the resultant force at the specified suit-wearer interface region.
For a musculoskeletal system with additional contacts, solving the interaction forces is an indeterminate problem.
AnyBody solves the problem by casting it as an optimization problem, with the cost function set as the norm of muscle and contact forces, and with the constraints enforcing muscle forces to be pulling and contact forces as pushing.

\section{EXPERIMENTAL VALIDATION}
\label{sec:benchmark}

The test data collected during the human subject tests were imported and directly processed in MATLAB.

This section reports the experimental validation results of the proposed simulation modeling framework.

\subsection{Setup of Subject and EOD Suit}

\noindent \textbf{Human subject:}
This study is approved by the Institutional Review Board (IRB) of the University of Massachusetts Lowell (\#19-023).
In the pilot testing, one healthy human subject (31 years old, 169 cm, and 60 kg) was recruited.

\begin{figure}[t]
    \centering
    \includegraphics[width=0.8\linewidth]{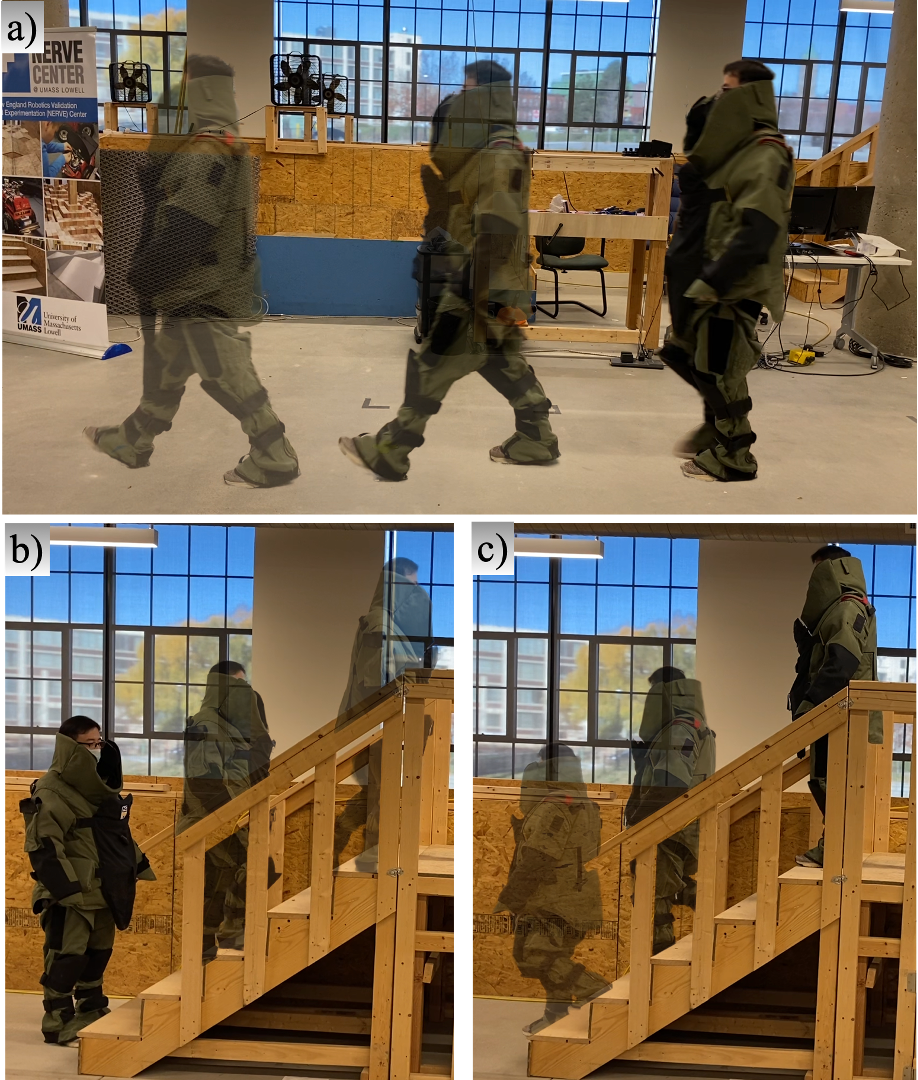}
    \caption{Time lapse figures of the three types of subject movements: a) walking on the flat ground, b) walking upstairs, and c) walking downstairs.}
    \label{fig:experiment_course}
\end{figure}

\noindent \textbf{Movement types:}
The pilot subject testing included three movement types, which were flat-ground walking, walking upstairs, and walking downstairs (see Fig.~\ref{fig:experiment_course}).
The distance of walking on the flat terrain was about 6.4 m.
The total height of the staircase with five flights was approximately 0.8 m.
Three trials were tested with each movement type.
During each trial, the movement sequence in the temporal order was quite standing, walking (on ground or stairs), and quite standing.

\subsection{Setup of Human-Suit Model in AnyBody}

In this study, we focus on validating the suit model in predicting the interface load at the subject's shoulders because shoulders have been reported as one of the body segments that are subject to significant discomfort during common suited movements~\cite{deane2021pressure}.
In AnyBody (Version 7.2), sixty contact points were defined to be evenly distributed within the top portion of each shoulder to ensure an accurate computation of the interface load without inducing an overly high computational load.

\subsection{Setup of Movement Sensors and Kalman Filter}

In this experiment, the in-suit motions of the human subject were measured by the APDM~\cite{APDM} inertial motion-capture system.
The system comprises a suite of conpact, light-weight inertial measurement units (IMUs) that can be worn on the subject (see Fig. \ref{fig:IMUs placement}-a).
The system processes the raw data returned by the IMUs to produce the estimated joint angles of the subject as well as the orientation of each IMU in the world frame. 
The APDM sensors return data at a rate of 128 Hz (i.e., the sampling period $\Delta t$ is $0.0078$ s), and its inaccuracy of base orientation measurement is 2.8$^\circ$.

Table II lists the noise standard deviations (SD) for the Kalman filter.
The values are tuned based on the nominal noise levels provided by the sensors' manufactures for ensuring a reasonable convergence rate and final accuracy.
Although the human model in AnyBoby has 214 joints, the Kalman filter only utilized the hip, knee, and ankle joints to estimate the global-position of the human model.

\subsection{Setup of Pressure Sensors at Shoulder-Suit Interface}

During all experiments, the human subject wore APDM IMUs and EOD suit together with pressure sensors (see Fig.~\ref{fig:IMUs placement}).
The pressure sensors were used to verify the interface load produced by the proposed modeling approach.

\noindent \textbf{Pressure sensor selection:}
Pressures sensors developed by Novel Electronics Inc. were utilized to obtain the interface load at the top portion of the subject's left and right shoulders.
We tested both Pliance and insole Pedar sensors, and then chose to use the Pedar sensors instead of the Pliance sensors because of their higher accuracy in obtaining static and dynamic pressure measurements at the shoulder areas.
This is essentially due to Pedar sensors' concentrated measuring surfaces and more robust measurement range for highly concentrated moving loads.
The Pedar sensor system contains two sensors. 
Each sensor covers an area of $70 \times 160$ mm${}^2$, consists of $99$ sensing units with a resolution of $5$ kPa, and transmits data via Bluetooth.

\noindent \textbf{Pressure sensor placement:}
In this study, the interfacial dynamic loads between the suit and individual shoulders were collected.
 The left Pedar sensor was set in between the left pectoral and trapezius region (shoulder composition of clavicle and acromioclavicular joint), with the right sensor in the same region on the right-handed side~\cite{deane2021pressure}.
 The cables of the sensors were attached using a Velcro strap onto the outside of the EOD suit after the suit was worn by the human subject(s)~\cite{deane2021pressure}.
 The sensors were prevented from physically shifting during the tests through shoulder straps and kinesiology tapes applied directly on the subject's skin (see Fig.~\ref{fig:IMUs placement}).
 Individual sensors were checked to ensure no shifting occurred during the placement of the EOD suit throughout the body.
 During the active use of the EOD suit, the overall weight of the suit is distributed not only between the two shoulder regions, but some of the pressure is taken up by the chest, arms, torso, and back~\cite{deane2021pressure}.

\begin{table}[t]
\centering
\caption{\small{Noise standard deviation (SD) for Kalman filter}} 
\small
\begin{tabular}{ P{4.5cm}|P{1.8cm} }
\hline
\hline
\centering
Measurement type & Noise SD \\
\hline
 \centering Linear acceleration (m/s${}^2$) &  0.2 \\
 \centering  Support foot position (m) & $10^{-5}$ \\
 \centering  Swing foot position (m) & $10^8$ \\
 \centering  Joint angles  ($^\circ$)& 10 \\
\hline
\end{tabular}
\end{table}

\noindent \textbf{Pressure sensor calibration and re-zeroing:}
To ensure measurement accuracy and repeatability during dynamic human subject movements, the pressure sensors were carefully calibrated using the Trublu calibration device (developed by Novel Electronics Inc.). 
The device uniformly pressurizes the Pedar sensors to the maximum amount the sensors can withhold through several incremental steps. 
To remove the nonzero sensor reading caused by the pressure applied by the sensor anchoring mechanism (i.e., straps and tapes), the suit was taken off of the subject (with the anchoring mechanism still on) every three movement trials to re-zero the reading.

\noindent \textbf{Interface load computation in AnyBody and through pressure sensing:}
In AnyBody, we compute the resultant force from the shoulder area at each time step by directly summing the projections of the individual contact forces along the normal direction of the contact area.
This approximation is reasonably accurate because the tangential forces are less than 10\% of the normal forces in magnitude.
In experiments, the proprietary software of the Pedar pressure sensing system sums all the forces returned by the sensing units to provide the resultant force at each time step.

\subsection{Validation of Simulated Shoulder-Suit Interface Loads}

\subsubsection{{Results of Shoulder-Suit Interface Loads Obtained through Experimental Pressure Sensing}}
To evaluate the accuracy of the proposed modeling approach in reflecting the shoulder-suit interface loads, we first used experimental pressure sensing to obtain the relatively accurate approximations of the true interface loads.
As the output of the pressure sensor is the resultant rather than the pressure distribution, we use the resultant from both the experiment and AnyBody simulation to validate the simulation results.

\begin{figure}[t]
    \centering
    \includegraphics[width=1\linewidth]{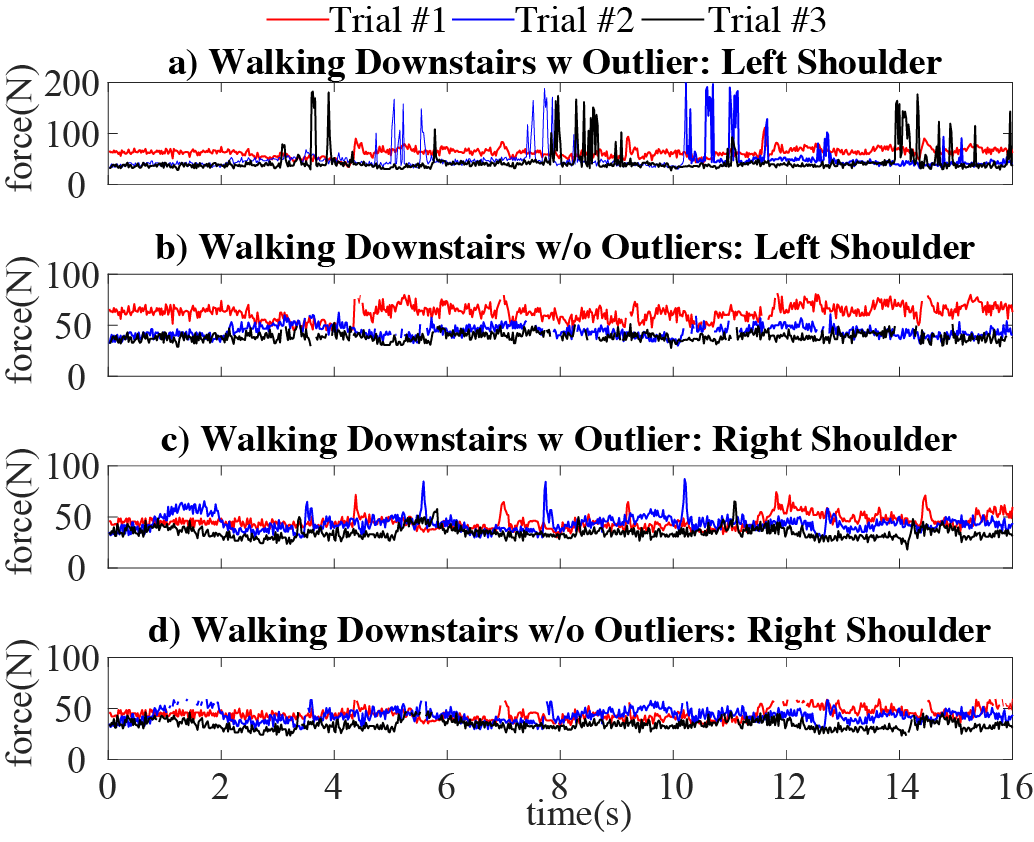}
        \vspace{-0.3 in}
    \caption{Interfacial shoulder loads when walking downstairs: 
    a) left-shoulder with outliers;
    b) left-shoulder without outliers; 
    c) right-shoulder with outliers; and
    d) Right-shoulder without outliers.}
    \label{fig:shoulder_load_outliers}
\end{figure}

\begin{figure}[t]
    \centering
    \includegraphics[width=1\linewidth]{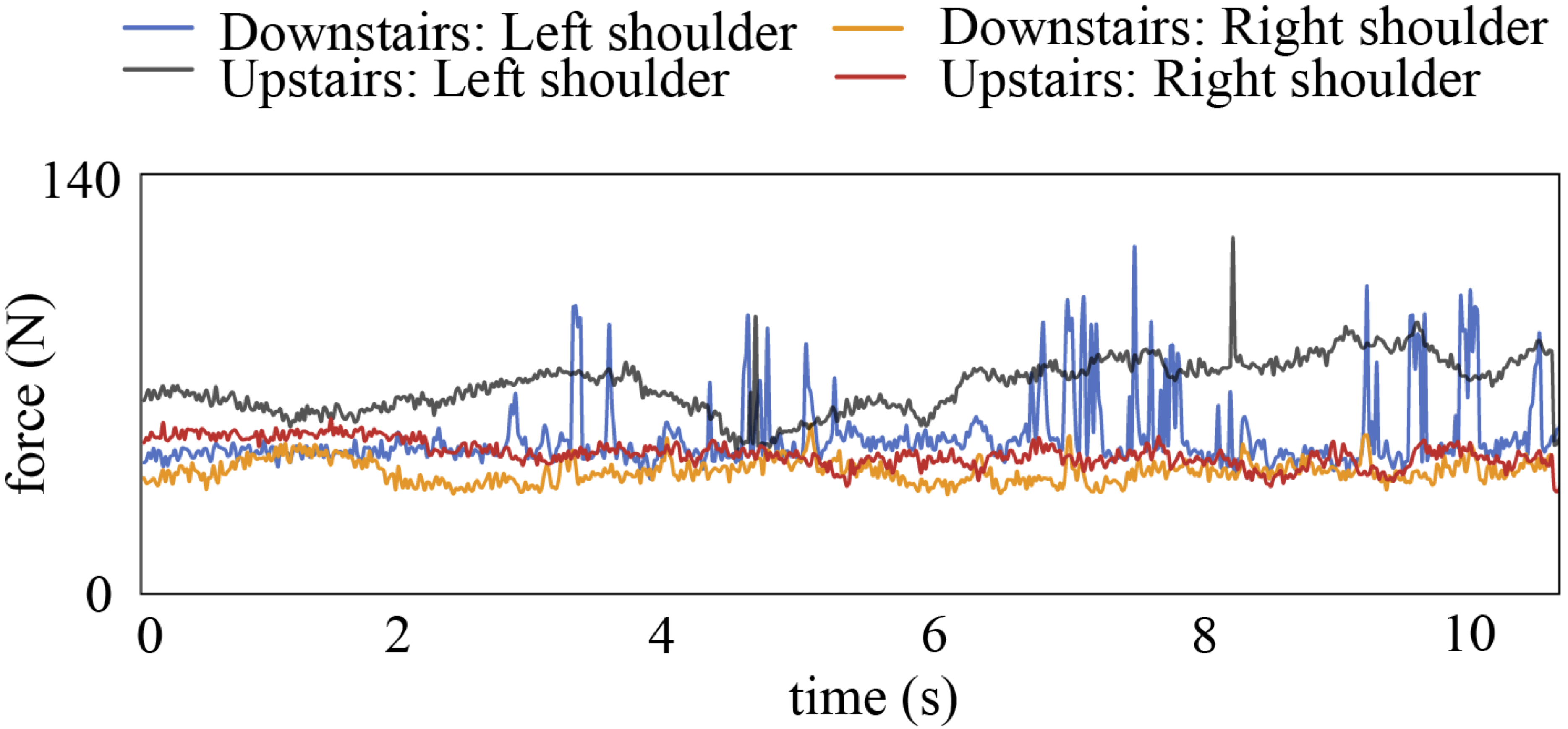}
    \vspace{-0.3 in}
    \caption{Average interface forces (with outliers retained) at the shoulder-suit contact regions for all trials of upstairs and downstairs walking.}
    \label{fig:shoulder_comparioson}
\end{figure}

Figure~\ref{fig:shoulder_load_outliers} a) and c) display the interface loads at the subject's shoulders obtained via pressure sensing.
These two plots show relatively significant spikes (i.e, outliers) that intermittently appear within short periods of time.
Furthermore, in Fig.~\ref{fig:shoulder_comparioson}, the average interface forces of all trials (with outliers retained) were graphed for left and right shoulders and for upstairs and downstairs walking.
The figure displays that the right shoulders from both (upstairs and downstairs walking) experiments are closely correlated to each other, whereas the left shoulder data exhibited more noise.

The outliers correspond to the unexpected spikes in the experimental data due to sudden impacts detected by the sensors.
Such a sudden impact can be an impact between the shoulder and the suit induced by foot-landing events.
Other causes of the faulty data could be the non-symmetrical dimensions of the left and right shoulder regions as well as sensor bending uncorrelated with the physical tasks.

\begin{figure}[t]
    \centering
    \includegraphics[width=0.95\linewidth]{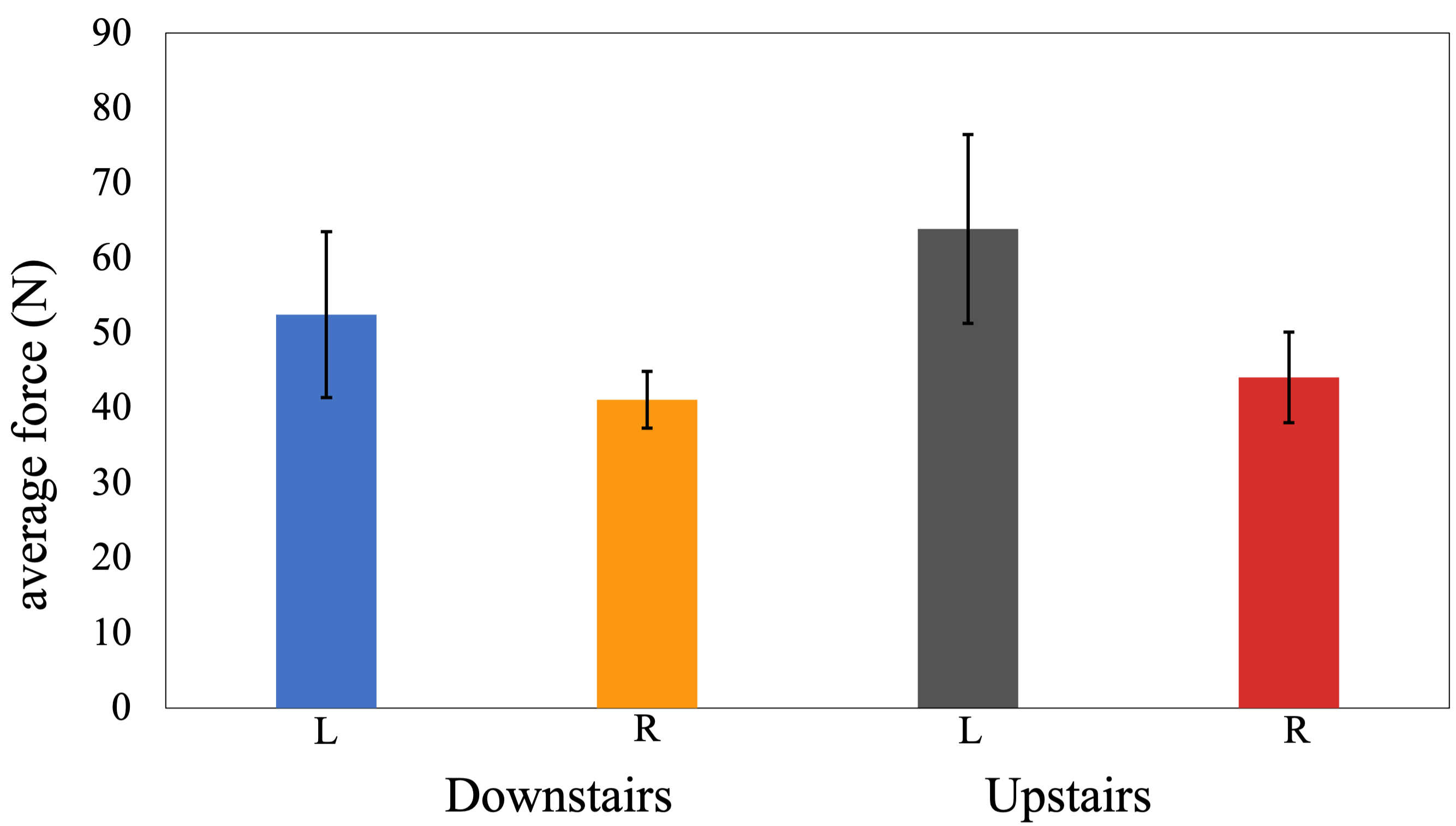}
    \vspace{-0.15 in}
    \caption{The average values and the ranges of variations for all trials of upstairs and downstairs walking at left (L) and right (R) shoulders.}
    \label{fig:sd_upstairs_downstairs}
\end{figure}

These outliers were consistently removed using the Interquartile range rule as part of the data analysis and interpretation processing~\cite{fundamental_of_data_visualization}.
Data was filtered by determining where 95 percent of the results fell between and using standard deviation to differentiate the accuracy.
All force readings exceeding three-halves the mean ($\mu\pm 1.5\sigma$), or data that fell outside the $95^{th}$ percentile, was considered noise.

Figure~\ref{fig:shoulder_load_outliers} b) and ~\ref{fig:shoulder_load_outliers} d) show the interface loads at the shoulders after outlier removal. 
These plots indicate that the pressure distribution along the shoulder projected from the EOD suit was consistent.
Specifically, Trials $\#2$ and $\#3$ were discerned to be the most congruent with each other.

In Fig.~\ref{fig:sd_upstairs_downstairs}, the bars (blue, yellow, grey, and red) indicate the average values of the pressure data.
The upper and the lower whiskers, respectively, indicate the maximum and the minimum pressure values.
The figures exhibit that the variation of the standard deviation was small for the right shoulders in both walking upstairs and downstairs while left-shoulder standard deviation experienced more variability.
This indicates the results for the right shoulder were relatively more accurate.

\begin{figure}[t]
    \centering
    \includegraphics[width=1\linewidth]{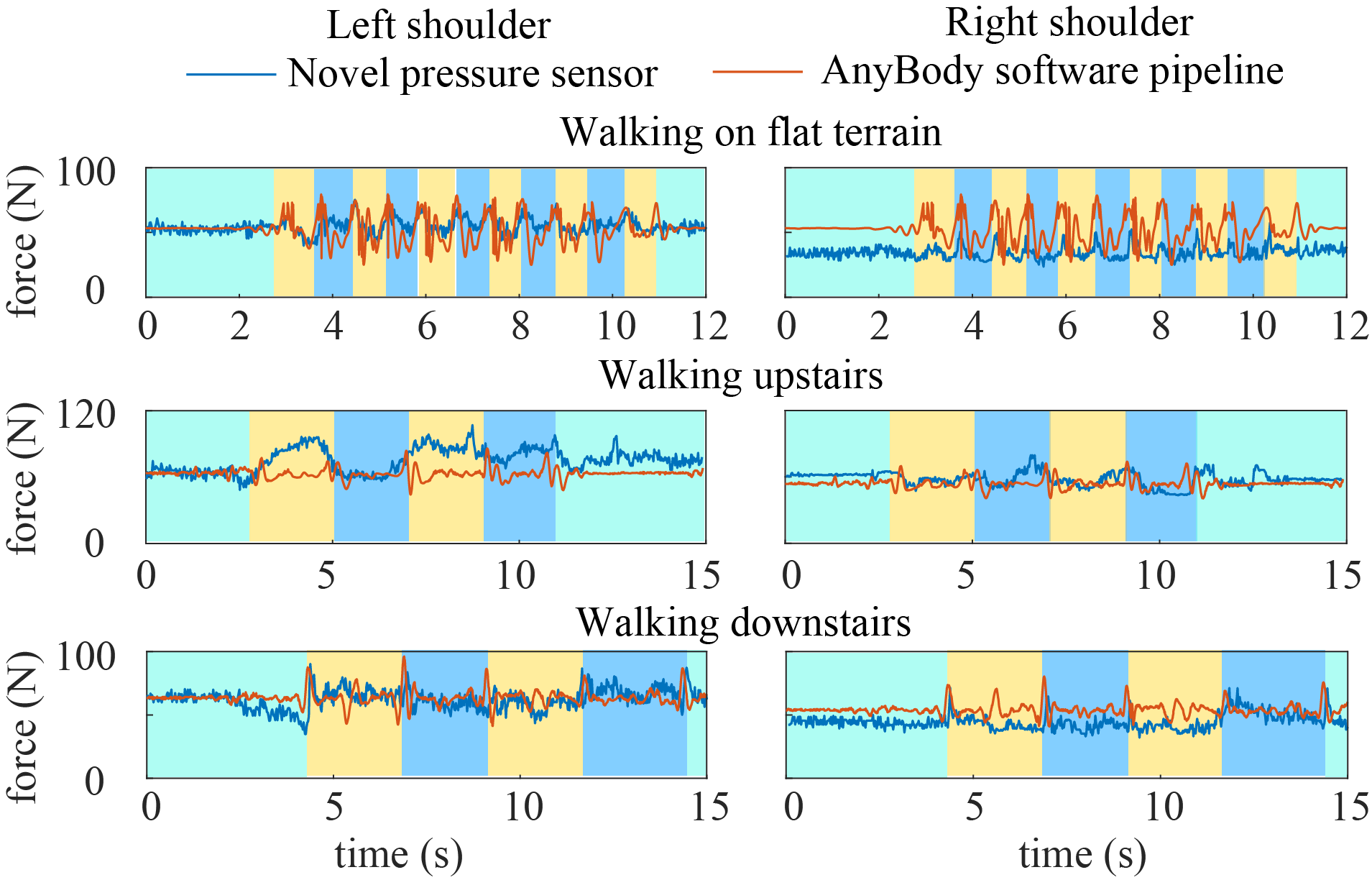}
    \vspace{-0.25 in}
    \caption{Shoulder-suit interaction forces computed based on AnyBody inverse dynamics analysis for three subject motions.
    The green shaded area indicates that the human subject stands quietly with two feet on the ground.
    The yellow (and blues) shaded areas correspond to the periods during which only the right (and left) foot contacts the ground.}
    \label{fig:human-suit interaction}
\end{figure}

\begin{table}[t]
\centering
\caption{\small{RMS errors of all movement trials.}}
\small
\begin{tabular}{ P{2.55cm}|P{2.25cm}|P{2.45cm} }
\hline
\hline
Motion type & Left shoulder (N) & Right shoulder (N) \\
\hline
Walking on ground &  19.56 & 25.96 \\
Walking upstairs & 19.32 & 8.22 \\
Walking downstairs & 8.22 & 10.16 \\
\hline
\end{tabular}
\end{table}

\subsubsection{Results of Simulated Shoulder-Suit Interface Loads}

Figure~\ref{fig:human-suit interaction} displays the interface loads obtained based on the proposed modeling framework and pressure sensing for three movement types (i.e., walking on the flat terrain, upstairs, and downstairs).
The figure indicates that the average values and overall trends of the simulated and experimental loads are relatively close.
This is confirmed by the RMS errors of all trials for the three motions as given in Table IV. 

However, the errors between the simulated and experimental interface loads appear to be significant at the left shoulder during upstairs walking. 
This large error could be caused by the relatively inaccurate reading of the pressure sensors at the left shoulder during the trial, as discussed in Section III-E-1). 
This is also based on the observations of the figure that : a) the experimental forces at the left and right shoulders during upstairs walking show relatively large discrepancies and b) the simulated and experimental forces at the right shoulder show relatively close correspondence.

Moreover, for flat terrain walking, the experimental and simulated forces at the right shoulder have an offset of approximately 30 N while there is no obvious offset at the left shoulder. 
In particular, the experimental interaction forces at the right shoulder has a nearly constant offset compared with that at the left shoulder.
This implies the interface force error between simulations and experiments for the right shoulder could be caused by the relatively inaccurate pressure sensing for that displayed trial.

\subsection{Discussion of Validation Results}
\label{sec:discussion}

From inverse dynamics results, we noticed that although the trends match well, the magnitudes suffer from the discrepancy between experimental data and simulation data. 
We have investigated this issue and found a few potential causes for this issue.
We have found that the pressure sensor on the shoulder only covers a portion of the shoulder, and thus the interaction outside of the sensor coverage cannot be detected.
Also, the pressure sensor is only capable of detecting the normal force between shoulder and EOD suit, but the shear force cannot be detected.
Finally, there exists a geometry discrepancy between the actual and modeled suits which may cause inaccurate force computation. 

\section{Conclusion}
\label{sec: conclusion}
This paper has introduced a simulation-based modeling framework that computes the interaction forces between an EOD suit and its human wearer during different mobility tasks.
The framework comprised three main components, which are: a) 3-D modeling of the suit for accurately and efficiently capturing its physical properties, b) movement state estimation for producing the wearer's in-suit motions based on data returned by wearable inertial motion-capture sensors, and c) inverse dynamics analysis based on the simulated human-suit model and estimated human movement.
The effectiveness of the framework in producing accurate human-suit interaction loads during different wearer movements was experimentally validated through the comparison with the loads measured by commercial pressure sensors.

To improve the accuracy of the interface loads produced by the proposed modeling framework, we will increase the fidelity of the proposed suit model by including the suit's helmet and groin components, and will validate the framework through movement experiments with a larger number of human subjects and for an even wider variety of human movements.
To obtain more complete and reliable ground-truth data for result validation, we will utilize pressure sensors with customized shapes to measure the contact regions at multiple critical locations on a wearer (e.g., shoulders, thighs, and back) and to ensure sufficient sensor coverage at those locations are experimentally measured. 
More importantly, we will investigate how the Anybody simulation results could help improve EOD suit designs.

\section{Acknowledgements}
This project is sponsored by the Department of
the Army, U.S. Army DEVCOM Soldier Center (SC).
Distribution Statement:
Approved for public release;
distribution is unlimited (PAO \#: PR2022\_14825).
Thank you to the NERVE Center, TRACE Lab, and SDASL at UMass Lowell for providing the test course and research assistance. 

\bibliography{Reference1}

\bibliographystyle{ieeetr}

\end{document}